\documentclass[twocolumn,showpacs,preprintnumbers,amsmath,amssymb,superscriptaddress,10pt,aps,prl]{revtex4-1}
\usepackage{epsfig,amsopn}
\usepackage{graphicx}
\usepackage{epstopdf}
\usepackage{sidecap}
\usepackage{subcaption}  
\usepackage{amsmath,amssymb}
\usepackage{amsthm}
\usepackage{enumerate}
\usepackage{color}

\begin{document}
\title{Magnetic Flux Periodicity in Second Order Topological Superconductors}
\author{Suman Jyoti De}
\affiliation{Harish-Chandra Research Institute, HBNI, Chhatnag Road, Jhunsi, Allahabad 211 019, India.}
\author{Udit Khanna}
\affiliation{Raymond and Beverly Sackler School of Physics and Astronomy, Tel-Aviv University, Tel Aviv, 6997801, Israel}
\author{Sumathi Rao}
\affiliation{Harish-Chandra Research Institute, HBNI, Chhatnag Road, Jhunsi, Allahabad 211 019, India.}

\begin{abstract}
  The magnetic flux periodicity of $\frac{hc}{2e}$ is a well known manifestation 
  of Cooper pairing in typical s-wave superconductors. In this paper we theoretically 
  show that the flux periodicity of a two-dimensional second-order topological superconductor, 
  which features zero-energy Majorana modes localized at the corners of the sample, is 
  $\frac{hc}{e}$ instead. We further show that the periodicity changes back to $\frac{hc}{2e}$ 
  at the transition to a topologically trivial superconductor, where the Majorana 
  modes hybridize with the bulk states, demonstrating that the doubling of periodicity 
  is a manifestation of the non-trivial topology of the state. 
\end{abstract}

\maketitle

\section{Introduction}

Topological insulators and superconductors are examples of symmetry-protected
topological phases (SPTs) which feature a gapped bulk spectrum with gapless modes
localized at the boundaries~\cite{HasanKaneRMP,QiZhangRMP,BernevigBook}. Free 
fermion SPTs with internal symmetries, such as charge-conjugation or 
time-reversal, can be completely classified in any dimension on the basis 
of  which of them are  present in the system~\cite{Ludwig2009,Kitaev2009,Ludwig2010}. 
Recent years have seen the advent of new classes of SPTs with spatial or crystalline
symmetries. These systems have a much richer connection between the topological
properties of the bulk and the states at the boundary. While crystalline topological
insulators~\cite{Dziawa2012,Tanaka2012} are analogous to the standard SPTs 
with a gapped bulk and gapless boundaries, higher-order topological 
phases~\cite{Hughes2014,JanSlager2015,Hughes2017a,Hughes2017b,Brouwer2017,
Kruthoff2017,Neupert2018,Fang2017,Fu2017,Watanabe2018,Brouwer2018,Khalaf2018,Queiroz2018, 
Brouwer2019,Seshadri2019,Ghorashi2019,Agarwala2019,Nag2019,Kudo2019,Tiwari2019,Ghorashi2019b}
have a gapped bulk with boundaries which are themselves topologically non-trivial. 
The $n^{\text{th}}$ order topological phase in $d$ dimensions has gapless 
modes at its $(d-n)$-dimensional boundary. 

Higher order topological insulators are best understood in the framework of 
the dipole moment theory of SPTs~\cite{Hughes2017a,Hughes2017b}. In this theory, the dipole moment 
of a crystal is defined in terms of a Berry's phase and the quantization (in 
presence of certain symmetries) of this moment leads to topological insulators
with boundary charges. This idea can be generalized to higher multipole moments, 
such as the quadrupole and octopole moments defined in terms of nested Wilson 
loops~\cite{NeupertBook}. Again in presence of certain crystalline symmetries,
these moments are quantized and lead to higher-order topological insulators with
boundary charges at the hinges or the corners. A topological invariant characterizing
the higher-order TIs can be  obtained from the Wilson loops in the same way as one
obtains the topological invariant from the Floquet operator, familiar in the context
of periodically driven systems~\cite{Hughes2017a,Hughes2017b,Franca2018}.

Second order topological superconductors, in analogy with higher order topological 
insulators,  can be written in terms of a mean field Boguliobov-de-Gennes Hamiltonian 
describing the bulk gapped  $d$ dimensional superconductor with gapless $d-2$ edge
states, instead of $d-1$ dimensional edge states as for usual topological superconductors.

The standard one dimensional topological superconductors~\cite{Qi2011,Alicea2012,Beenakker2013,Ramon2017} host edge  
Majorana modes, which are expected to obey non-abelian statistics, and hence be 
relevant for quantum computation. They have been shown\cite{Kitaev2001,Sengupta2004}  to exhibit the fractional 
Josephson effect, where the current-phase relation has a $4\pi$, 
rather than a $2\pi$ periodicity.  Other attempts to probe the topological order includes using non-linear Coulomb blockade
using a superconducting nanoring\cite{Zocher2012,Zocher2013}, tunneling spectroscopy\cite{Wimmer2011, Leijnse2011}  and transport
experiments\cite{Tewari2012}. There have also been proposals
to measure the flux periodicity in a  ring geometry with either a single\cite{Tripathi2016} or multiple\cite{Rubbert2016, Chiu2018,Sau2019} Majorana modes. 
An alternate system is the chiral $p$-wave superconductor, 
which is predicted to occur when the chiral edge modes of a quantum anomalous Hall insulator turn
superconducting via the proximity effect\cite{Qi2010} and lead to chiral Majorana states and there has been
some experimental evidence\cite{He2017,Shen2018} for these modes.
However, it has proved remarkably difficult to 
unambiguously prove the existence of these Majorana modes.

In this paper, we focus on a two-dimensional second order topological superconductor which 
hosts zero energy Majorana modes localized at the corners of the sample. We study the flux 
periodicity of the superconducting state after introducing a vortex in the center of the sample. 
The vortex makes the geometry multiply connected and thus the superconducting phase 
winds around the vortex in a non-trivial way. To take this into account, we compute the ground
state of the mean-field BdG Hamiltonian self-consistently at each value of the flux. This 
self-consistent calculation shows us that the flux periodicity of the second order topological
superconductor is $\frac{hc}{e}$ instead of $\frac{hc}{2e}$ as expected for a superconductor.
To probe the origin of this period doubling, we compute the flux periodicity while varying
a parameter in the Hamiltonian which drives the system into a topologically trivial phase. 
Interestingly we find that the flux periodicity changes to $\frac{hc}{2e}$ across this transition,
proving that the change in flux periodicity is related to the topologically nontrivial nature
of the state.

The plan of the paper is as follows. In section II, we introduce the model, originally studied 
in Ref.~\cite{Wang2018}, with $p_x+ip_y$ pairing in a doped Dirac semimetal with two mirror symmetries
- i.e., four mirror symmetric Dirac nodes. We will then show using a concrete pairing mechanism 
that a second order topological superconductor $TSC_2$ can be self-consistently realised in such 
a model, with four Majorana corner modes. We will then introduce a vortex through the centre in 
an annulus  geometry  (a square annulus in the lattice model) and obtain the self-consistent 
solutions of the superconducting order parameter, as a function of the parameters of the theory. 
In Sec III, we will  study the energy levels  and the  circulating current due to the insertion 
of the vortex and show that the flux periodicity changes from $hc/e$ to  $hc/2e$ as a  tunable parameter
in the model is changed. Further tuning of the parameter to bring the system into the 
metallic regime, changes the flux periodicity back to $hc/e$
as expected for an Aharonov-Bohm ring.  Finally, in Sec. IV, we end with discussions and conclusions.

\section{Model and vortex  introduction}
\subsection{A. Second order topological superconductor($TSC_2$)}

We start with the four band  Boguliobov- de Gennes model introduced in Ref.[\onlinecite{Wang2018}] with $H=\int d{\bf k} \Psi_{\bf k}^\dagger \mathcal{H}(\textbf{k}) \Psi_{\bf k}$,
\begin{align}
 \mathcal{H}(\textbf{k}) = (b_x + \lambda \cos(k_x))\tau_z \sigma_x + \lambda \cos(k_y) 1_\tau \sigma_y\nonumber  \\  + \Delta \sin(k_x) \tau_y \sigma_x + \Delta \sin(k_y) \tau_x \sigma_x , \label{eqone}
\end{align}
and  $\Psi_{\textbf{k}}^{\dagger} = (c_{\textbf{k}\uparrow}^{\dagger}, c_{\textbf{k}\downarrow}^{\dagger}, c_{-\textbf{k}\uparrow}, c_{-\textbf{k}\downarrow}).$
Here, $\sigma ( \tau )$ denote the operators acting on spin (Nambu) space respectively and $1_\tau$ represents the identity in the Nambu space. 
$\lambda$ denotes the hopping.
This model can be shown\cite{Wang2018} to describe a higher order  topological $p_x+ i p_y$ superconductor phase for a fixed $\Delta $,  when $|b_x/\lambda| < 1$,  with four Majorana modes localized at the 
four corners of the sample. This 
  Hamiltonian has a particle-hole symmetry,  with $\tau_x$ being the charge conjugation operator such that $ \tau_{x} \mathcal{H}(\textbf{k})^{T} \tau_{x}^{-1} = -\mathcal{H}(-\textbf{k}).$
  Provided that we choose the pairing terms to have $p_x+ip_y$ symmetry,  the model also has two mirror symmetries $M_x = \sigma_y\tau_y$ and $M_x = \sigma_y\tau_x$ such that $M_{x,y} \mathcal{H}({\bf k}) M_{x,y}^{-1} = H({\hat m}_{x,y} {\bf k})$ with the two
  mirror symmetries anti-commuting with each other.  More specifically, $M_x H(k_x,k_y)M_x^{-1} = H(-k_x,k_y)$ and similarly for $M_y$.
  
\begin{figure}[t]
\centerline{
\begin{subfigure}{4.0cm}
\caption{}
\vspace{0.05cm}
\includegraphics[width=3.8cm]{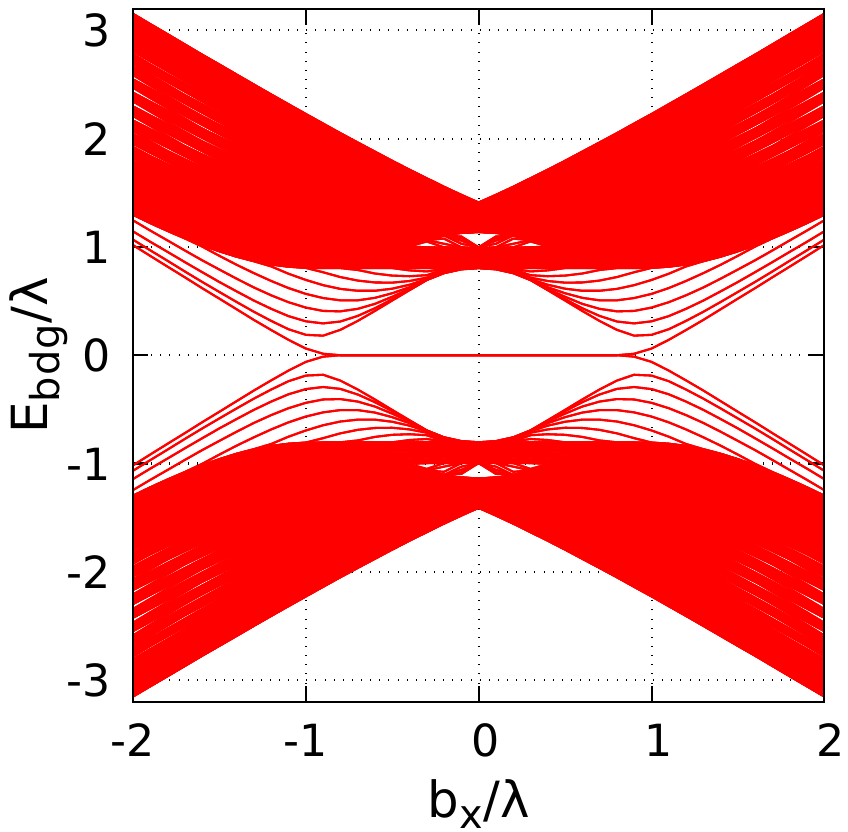}
\end{subfigure}
\hspace{0.25cm}
\begin{subfigure}{4.0cm}
\vspace{-0.15cm}
\caption{}
\vspace{0.03cm}
\includegraphics[width=3.93cm]{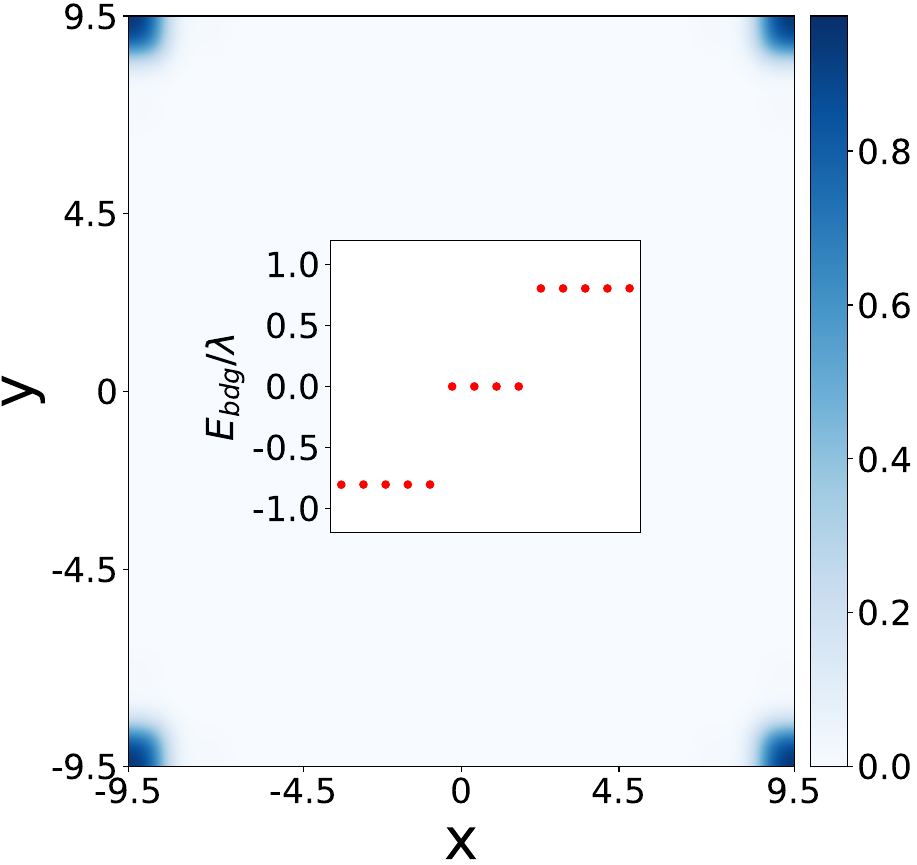}
\end{subfigure}
}
\caption{(a) Plot of the spectrum of the Hamiltonian in Eq.\ref{eqone}  for a  $20 \times 20$ lattice, for open boundary conditions in both directions,  with respect to the field $b_x$,  which clearly shows  the zero modes for $|b_x| < \lambda$. Parameter values are $ \lambda = 1 $ , $ \Delta/\lambda = 0.8 $ and lattice spacing $a=1$.
(b) Plot of the  electron densities which shows the four localized Majorana modes at the  four corners. The inset shows four zero 
energy modes clearly distinguishable from the bulk spectrum at $b_x=0$.}
\label{figone}
\end{figure}

\begin{figure}[hb]
\centerline{
\includegraphics[width=8.5cm,height=3.6cm]{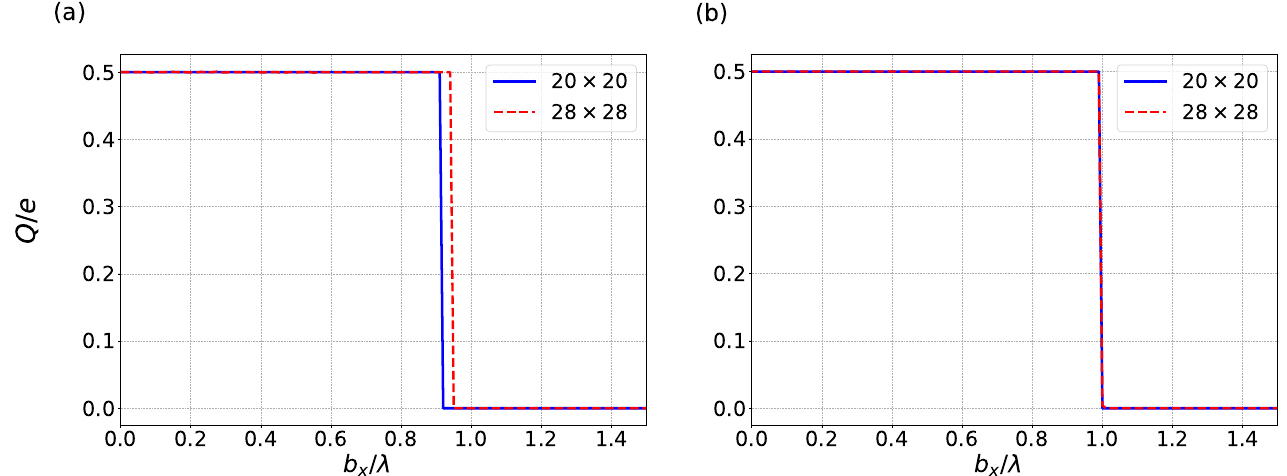}
}
\caption{The quadrupole moment($Q$) as a function of $b_x/\lambda$ for the Hamiltonian in Eq.\ref{eqone} with  $\lambda=1$, $\Delta/\lambda=0.8$ and lattice spacing $a=1$. (a) For open boundary conditions in  both $x$ and $y$ directions and  (b) for open boundary conditions in the  $y$ direction and periodic boundary conditions in the $x$ direction. This  shows that the  second order topological superconductor phase has $Q/e = 0.5$(modulo $1$) and the topologically trivial phase has $Q/e = 0.0$(modulo $1$). The phase transition from a topological to a non-topological superconducting phase occurs at $b_x/\lambda = 1.0$ in (b) showing that the minor deviation from unity in (a) here as well as in Fig.\ref{figone}(a) is a finite size effect.}
 \label{figtwo}
 \end{figure}

  This model has a gapped spectrum, but  as shown in Ref.[\onlinecite{Wang2018}], for $|b_x| < \lambda$, the model denotes a second order topological superconductor $TSC_2$ - i.e.,
  the edge states themselves are topological
  and have  gapless corner states.  Analogous to what was done for the model of higher order topological insulators in Ref.\onlinecite{Benalcazar2018}, we can plot
  the spectrum for open boundary conditions in both the  $x$ and $y$ directions, parametrically,  as a function of $b_x$, as shown in Fig.\ref{figone}.
  The spectrum in Fig.\ref{figone}(a) clearly shows the existence of zero modes for $|b_x/\lambda| < 1$, and in Fig.\ref{figone}(b), the densities clearly show 
   four localised modes at the four corners of the lattice.

A two dimensional quadrupole insulator can be characterized by  a quantized quadrupole moment($Q$)  as argued in Refs.\onlinecite{Hughes2017a,Hughes2018,Kang2018} and  
 the quadrupole moment is defined as a ground state $|\Phi_0\rangle$ expectation value of a many-body operator as follows -  
 \begin{align}
   Q=\frac{e}{2\pi} \mathrm{I} \mathrm{m}[\text{ln} (\langle\Phi_0|\hat{O}|\Phi_0\rangle)] \hspace{0.1cm} (\mbox{modulo 1}),\nonumber
 \\
   \hat{O}= \exp \bigg(2 \pi i \frac{1}{L_x L_y} \sum_{x,y} x y \, \hat{n}(x,y)\bigg) ,
  \label{eqtwo}
 \end{align}
where $(x,y)$ is the lattice site index, $\hat{n}(x,y)$ is the quasiparticle density at the site $(x,y)$ and $L_x,L_y$ are the lengths of the $2d$ system. By analogy,  we can define a similar quadrupole moment for the two dimensional topological superconductor, which has been plotted in Fig.\ref{figtwo}.  The quadrupole moment shows a sharp transition  from $Q=0.5$ to $Q=0$ at the value of $b_x$ where the model transitions from a topological superconductor into a normal superconductor.  This transition occurs close to $b_x/\lambda =1$.  This is also consistent with the disappearance of the corner Majorana modes at  $b_x = 1$ as seen in Fig.\ref{figone}. Here, and in Fig.\ref{figtwo}(a), the minor deviation from unity is a finite size effect.
Note, however that the density plotted in the figure is that of the Boguliobons, linear combinations of the particle and hole operators
obtained  by diagonalising the Hamiltonian in Eq.\ref{eqone}  with the pairing term  $\Delta$. This is discussed further in the next section, where
we compute the quadruple moment with a self-consistent pairing term.

  The normal state of this Hamiltonian (when $\Delta=0$) has four gapless mirror symmetric Dirac points, 
 and it can be shown that at finite chemical potential, regions of the Fermi surface with opposite momenta always
 have the same spin texture. Hence,  it is natural\cite{Wang2018} for a spin triplet  superconducting gap to be induced 
 by electronic interactions.
 The pairing potential $\Delta_{ij}$ with 
 the appropriate $p_x+ip_y$ symmetry can be derived from a mean field treatment of the pairing interaction 
\begin{align}
H_{int}  = \frac {V}{2}  \sum_{\langle i,j\rangle }   (n_{i\uparrow}n_{j\downarrow} + n_{i\downarrow}n_{j\uparrow})
\label{eqthree}
\end{align}
where the $\langle i,j\rangle$ denotes nearest neighbour sites.
 We will show in a later section that $\Delta_{ij}$ can be obtained self-consistently for our model on a square lattice.

\subsection{B. Vortex insertion in an annulus geometry}

The basic idea is that in a superconducting ring, the order parameter responds to a flux or vortex inserted through the ring. Just 
as current through a metallic ring is modulated by a $hc/e$ periodicity due to the Aharanov-Bohm effect, the current through a superconducting
ring is expected to be modulated as $hc/2e$\cite{Byers1961, Onsager1961}.  Although naively explained in terms of the Cooper pair condensates having a charge of $2e$,
the theoretical explanation is more subtle and comes from the degeneracy between two different classes of superconducting wave-functions at $\phi=0$ and at $\phi=\phi_0/2$.
The first class consists of  those wave-functions with pairing between the angular momentum states $\hbar k$ and 
$-\hbar k$ leading to Cooper pairs with $\hbar q =0$. All even values of $q$ can obtained from these wave-functions by gauge transformations.
The second class consists of   those wave-functions with pairing between $\hbar  k=$ and $\hbar (-k+1)$
leading to Cooper pairs with $\hbar q=1$, with again, all odd integer values of $q$ being related to these wave-functions by gauge transformations.
Both these classes of wave-functions turn out to have the same energy for flux $\phi = 0$ and $\phi=hc/2e$.
But more recently, the question of the flux periodicity has resurfaced in the context
of high $T_c$ $d$-wave superconductors\cite{Loder2008} where it was seen that the condensate reconstructs for half-integer flux quanta, and breaks the degeneracy
between the state at zero flux and the state at  half-integer flux. Thus, the periodicity changes  back to $hc/e$ as for normal metals.
Even for $s$-wave superconductors, it has been shown\cite{Loder2012} that for superconducting rings with diameter smaller than the coherence
length, the response due to magnetic flux is generally modulated as $hc/e$ periodic instead of $hc/2e$.

\begin{figure}[t]
\centerline{
\includegraphics[width=7.5cm,height=6.5cm]{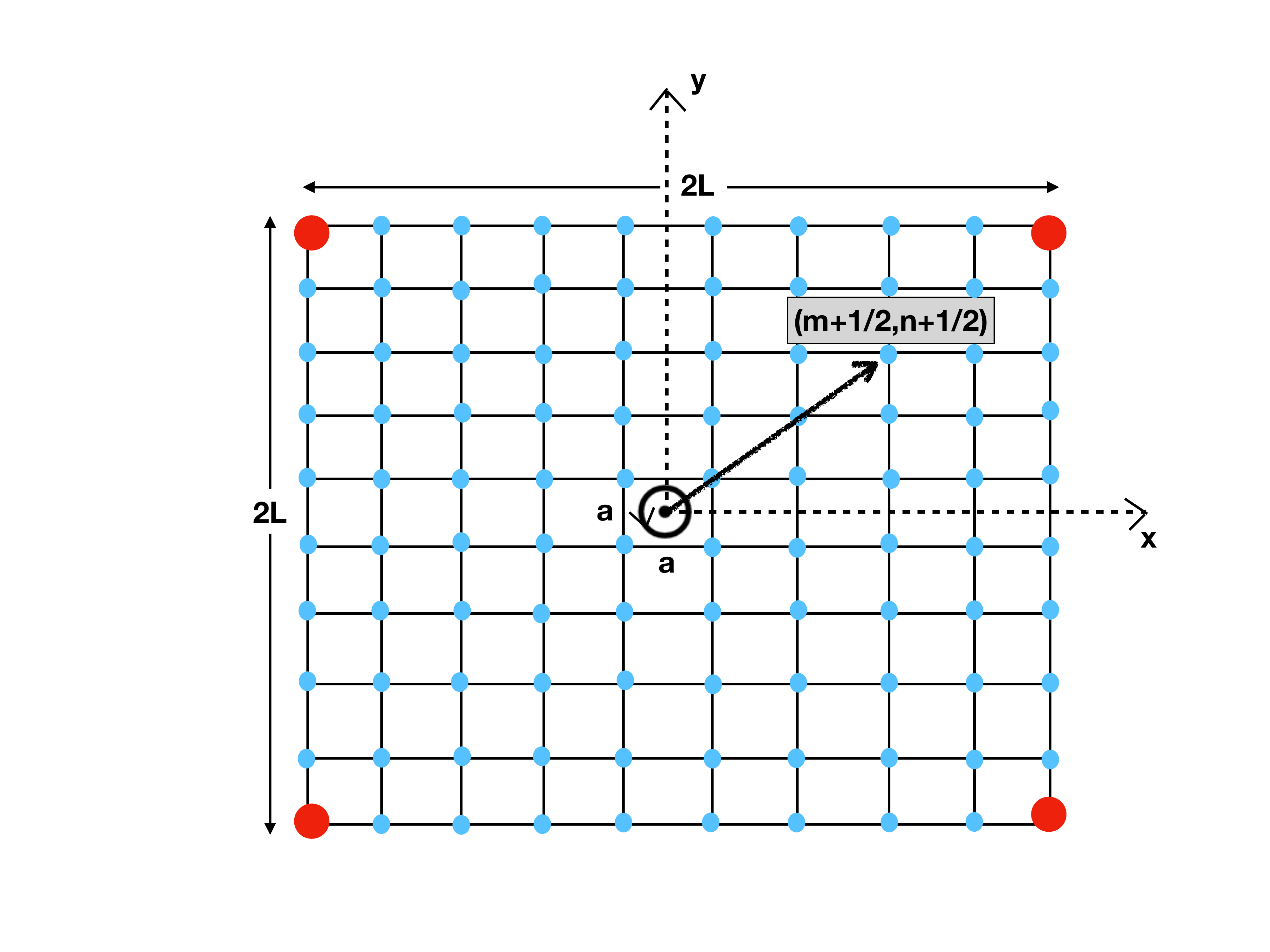}
}
\caption{Schematic diagram of flux insertion through the central plaquette ($a\times a$) of a square lattice ($2L\times 2L$) leading
to a ring geometry }
\label{fig:sys}
\end{figure}

Here we study the response in a $p_x+ip_y$ higher order topological superconductor `ring'.
We imagine adding an infinitely long solenoid (of infinitesimal radius) at the origin of a $2d$ sample so that there is no magnetic field crossing any of the sites, but a closed loop around the origin encloses a flux - thereby mimicking an annulus with flux through the hole. 
More specifically, we have a square geometry and assume that the lattice sites are located at $\textbf{r} = (m+1/2,n+1/2)$ where $m,n$ are integers from $-L$ to $L-1$. This ensures that the lattice sites are symmetrically located about the origin at $\textbf{r} = (0,0)$.
This has been shown in Fig.\ref{fig:sys}.

\begin{figure}[t]
\centerline{
\includegraphics[width=8cm,height=4.5cm]{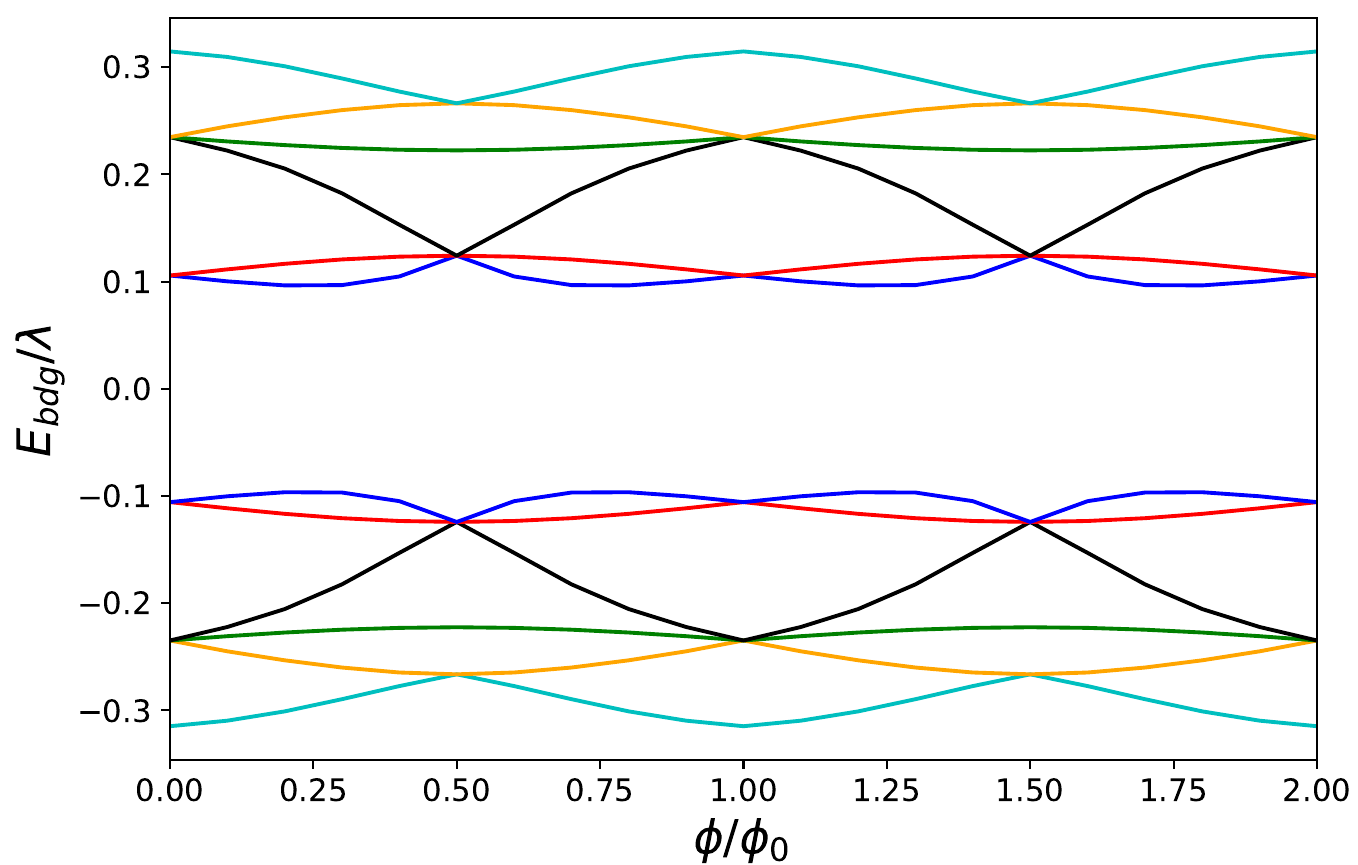}
}
\caption{Lowest three positive and negative eigenvalues
of the Hamiltonian  in Eq.\ref{eqone}   for a $20 \times 20$ lattice showing
their magnetic flux periodicity, as a function of the flux $\phi/\phi_0$ introduced at the origin. $\phi_0=hc/e$. The parameter values are  $\lambda=1,\Delta=0,b_x/\lambda=0$ and lattice spacing $a=1$.}
\label{figthree}
\end{figure}

\begin{figure}[h]
\centerline{
\includegraphics[width=7.9cm,height=4.5cm]{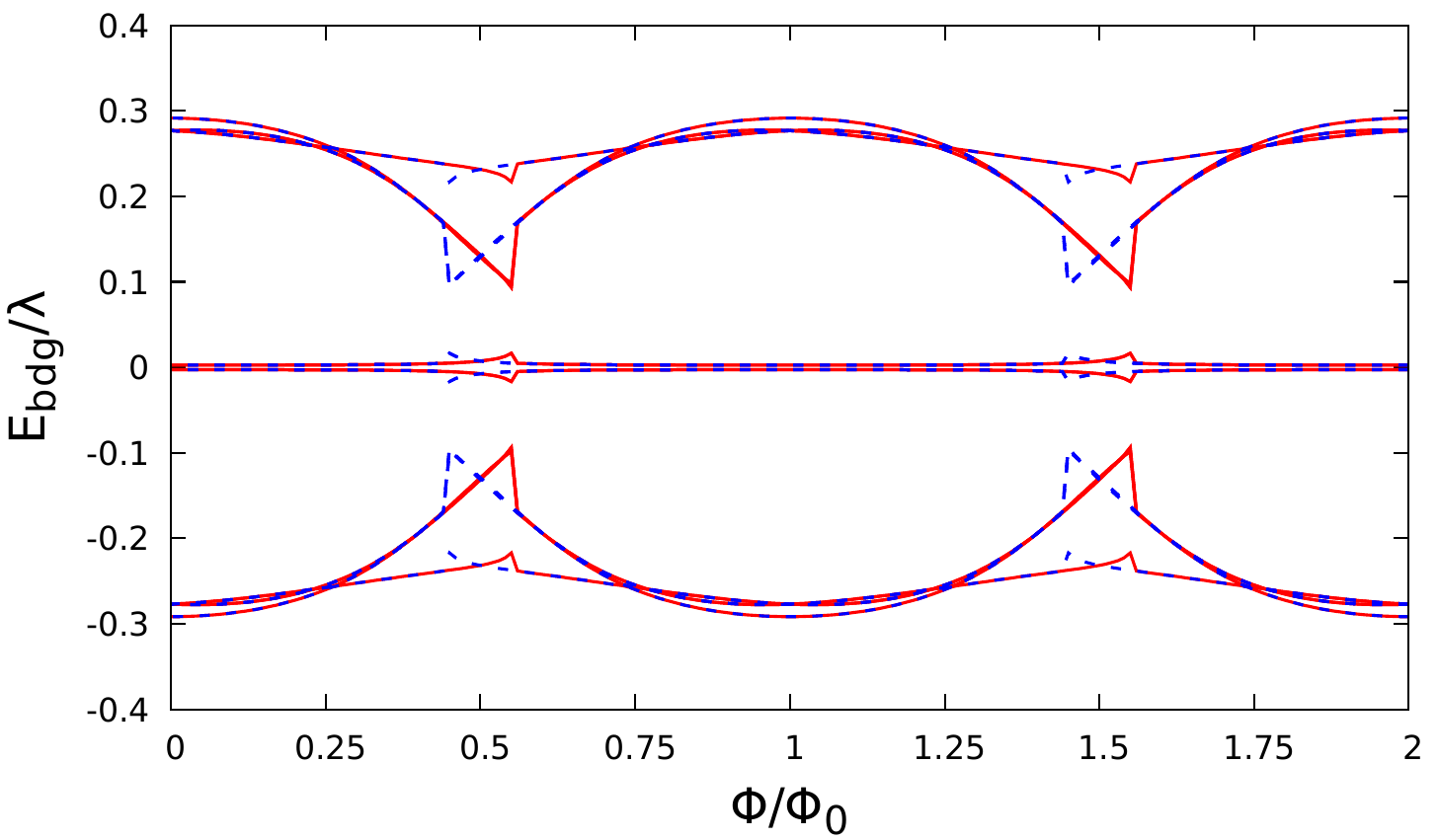}
}
\caption{Self-consistent lowest three positive and negative eigenvalues of the Hamiltonian in Eq.\ref{eqfive} for a $20 \times 20$ lattice, as a function of the flux $\phi/\phi_0$ introduced at the origin for $\lambda=1,V=3\lambda,b_x/\lambda=0$ and lattice spacing $a=1$.  Note the existence of two different self-consistent states (one is the  solid red line and other one is the dashed blue line) close to $\phi =\phi_0/2$.
}
\label{figfour}
\end{figure}

\begin{figure*}[ht]
\centerline{
\includegraphics[width=20.5cm,height=5.2cm]{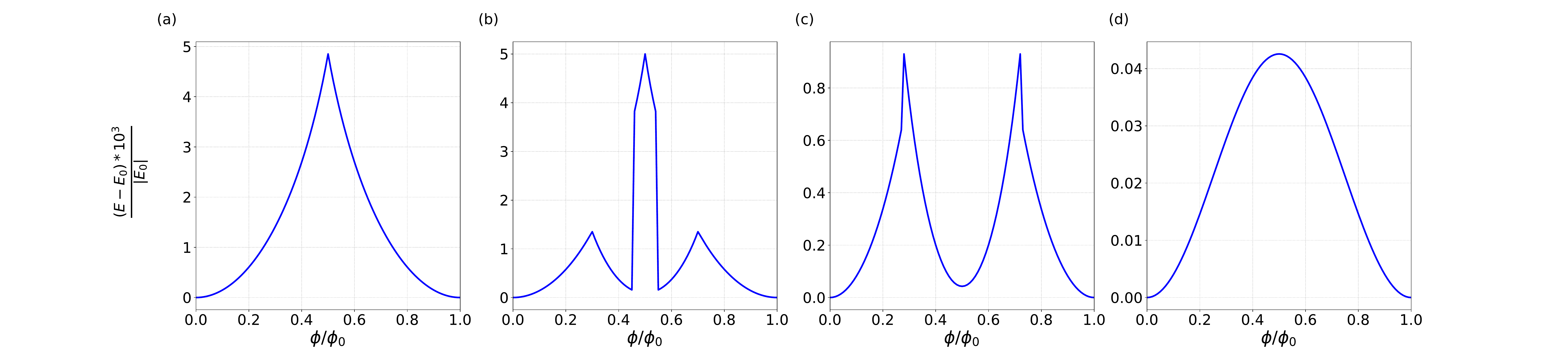}
}
\caption{Self-consistent total energy $((E(\phi/\phi_0)-E(0))*10^3)/|E(0)|$ as a  function of the flux $\phi/\phi_0$ through the center of a $20\times20$ square lattice having lattice spacing $a=1$ for the Hamiltonian in Eq.\ref{eqfive}. The parameters chosen are  $\lambda=1,V =3\lambda$ with (a) $b_x/\lambda=0.0$ (b) $b_x/\lambda=0.3$ (c) $b_x/\lambda=0.8$ and (d) $b_x/\lambda=1.5$. Note that the flux-periodicity changes from $\phi_0$ to $\phi_0/2$ and back  again to $\phi_0$, as  $b_x$ is tuned.} 
\label{figfive}
\end{figure*}

In the absence of superconductivity, a vortex  can be added to $\mathcal{H}$ through the standard Peierls substitution\cite{Peierls}.
Under this transformation, the kinetic terms change as follows - 
\begin{align}
 \psi_{\textbf{r}+\hat{\delta}}^{\dagger} \psi_{\textbf{r}} \rightarrow \psi_{\textbf{r}+\hat{\delta}}^{\dagger} \psi_{\textbf{r}} \exp\bigg(i \frac{e}{c\hbar} \int_{\textbf{r}}^{\textbf{r} + \hat{\delta}} d\textbf{r}^\prime \cdot \vec{A}(\textbf{r}^\prime)  \bigg) = \nonumber 
\\  \psi_{\textbf{r}+\hat{\delta}}^{\dagger} \psi_{\textbf{r}}  \exp\bigg(i \frac{\phi}{\phi_0} \int_{\textbf{r}}^{\textbf{r}+\hat{\delta}} d\textbf{r}^\prime \cdot \hat{\theta} \frac{1}{|\textbf{r}^\prime|}  \bigg)
\label{eqfour}
\end{align}
where $\phi_0 ={hc}/{e}$ , $\psi_{\textbf{r}}^{\dagger} = (c_{\textbf{r}\uparrow}^{\dagger},c_{\textbf{r}\downarrow}^{\dagger},c_{\textbf{r}\uparrow},c_{\textbf{r}\downarrow})$ and $\hat{\delta}$ is constrained only upto nearest neighbour in both $\hat{x}$ and $\hat{y}$ direction. 
Every bond of the lattice will clearly pick up a different phase due to the $\frac{\hat{\theta}}{|\textbf{r}^\prime|}$ in the integral, but at $\phi=n\phi_0$ (where $n$ is an integer), the total phase accumulated by an electron going through each plaquette around the origin is $2\pi n$. Therefore the system behaves as if there is no magnetic field at all. For illustration, the lowest
three positive and negative eigenvalues are shown as a function of the flux in Fig.\ref{figthree}, which show that the spectrum is $\phi_0=\frac{hc}{e}$ periodic as expected.

However, in the presence of pairing terms ($\Delta \neq 0$), adding a flux or creating a vortex at the center of system makes $\Delta$ 
position dependent and keeping it constant is no longer viable. We need to solve for $\Delta$ self-consistently, which is done in the next section.

\begin{figure*}[ht]
\centerline{
\includegraphics[width=20cm,height=5.2cm]{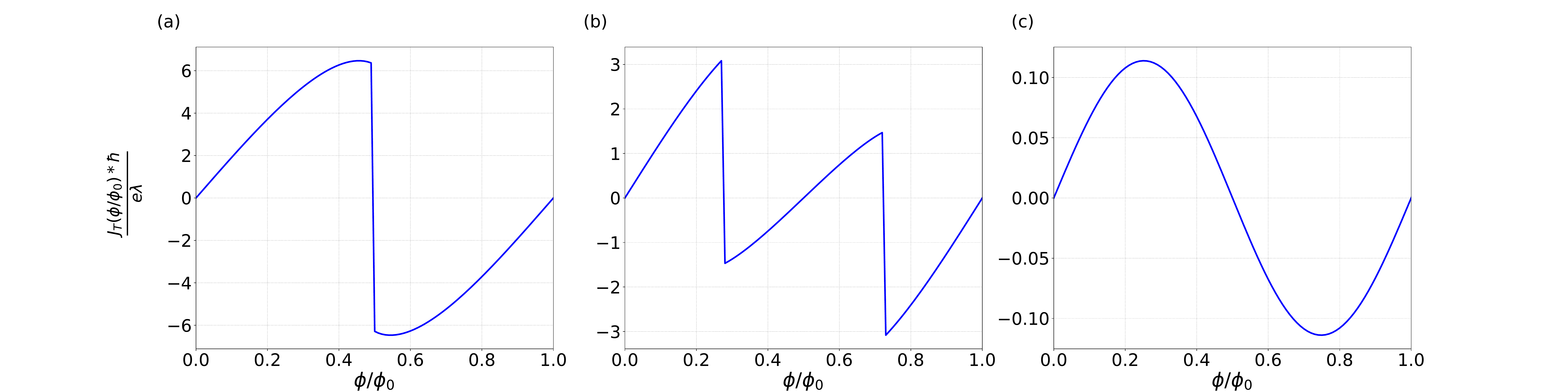}
}
\caption{Total circulating current $J_T(\phi/\phi_0)$ as function of flux $\phi/\phi_0$ through the center of $20\times20$ square lattice having lattice spacing $a=1$  for the Hamiltonian in Eq.\ref{eqfive}. $\lambda =1$ and $V=3\lambda$ as in Fig.\ref{figfive}.  Note that the flux periodicity is $\phi_0$ in (a)  with $b_x/\lambda=0.0$,  $\phi_0/2$ in (b) with $b_x/\lambda=0.7$  and  is again back to $\phi_0$ in (c) with $b_x/\lambda=1.5$.} 
\label{figsix}
\end{figure*}

\section{Self-consistent calculation and results}

We work with the Hamiltonian in Eq.\ref{eqone} in real space given by 
\begin{align}
\mathcal{H}=\sum_{\langle i,j \rangle} \bigg(\mathcal{H}_{ij}^0 c_{i\uparrow}^{\dagger} c_{j\downarrow} +h.c. \bigg) +\sum_{\langle i,j \rangle} \bigg(\Delta_{ij} c_{i\uparrow}^{\dagger} c_{j\downarrow}^{\dagger} + \nonumber
\\ \Delta_{ji}^* c_{i\downarrow} c_{j\uparrow} \bigg)\label{eqfive}
\end{align}
where  
$\mathcal{H}_{ij}^0 = \vec{t_h} \exp\big(i\phi_{ij}) + 2b_x\delta_{ij}$,  with $t_h^x=\lambda $  and $t_h^y=-i\lambda$,  are the hopping matrix elements between the nearest neighbour sites. Also $\phi_{ij}= \int_{\textbf{r}_{j}}^{\textbf{r}_{i}} \frac{e}{\hbar c} \big( \vec{A}(\textbf{r}^\prime) \cdot d\vec{\textbf{r}^\prime} \big)$ is the Peierls phase factor coming from $\vec{A}(\textbf{r})$, which  is the vector potential due to flux through the origin. 
The order parameter of the superconducting state $\Delta_{ij}$ is defined on the links between neighbouring sites with the  appropriate $p$-wave symmetry, i.e, we have 
\begin{align}
\Delta_{ij} = \frac{V}{2} \big[ \langle c_{i\uparrow} c_{j\downarrow}\rangle  + \langle c_{i\downarrow} c_{j\uparrow}\rangle  \big]
\end{align}
which is symmetric under the exchange of spins and anti-symmetric under the exchange of spatial  indices.
Here,  $V$ is the nearest-neighbour pairing interaction strength. 
Now following a Boguliobov rotation and in terms of the  new fermionic operators (the Boguliobons) $\gamma_n$
\begin{align}
c_{i\sigma} = \sum_n \bigg(u_{i\sigma}^n \gamma_n - (v_{i\sigma}^n)^* \gamma_n^{\dagger}\bigg) \label{eqseven}
\end{align}
we find that the Hamiltonian in Eq.\ref{eqfive} is given by 
\begin{align}
\mathcal{H} = E_g + \sum_n \big(E_n \gamma_n^{\dagger} \gamma_n\big), \hspace{0.2cm} E_n > 0 \label{eqeight}
\end{align}
with $E_g = -\sum_{i,\sigma,n} E_n |v_{i\sigma}^n|^2$ 
where the coefficients $(u^n)$ and $(v^n)$ satisfy  the equation
\begin{align}
\begin{pmatrix}
\mathcal{H}^0&-\Delta\\
-\Delta^{\dagger}&-{\mathcal{H}^0}^T\\
\end{pmatrix}
\begin{pmatrix}
u^n\\v^n
\end{pmatrix}
= E_n 
\begin{pmatrix}
u^n\\v^n
\end{pmatrix}.
\end{align}
Here $(u^n)$ and $(v^n)$ are $2N$ dimensional column vectors  ($N$ is the total number of lattice sites, and the $2N$ is because of spin) and  both $\mathcal{H}^0$ and $\Delta$ are $2N \times 2N$ matrices.
The order parameter $\Delta_{ij}(\phi,b_x) $ is then calculated self-consistently from the following equation 
\begin{align}
\Delta_{ij}=\frac{V}{4} \sum_n \bigg[u_{j\uparrow}^n (v_{i\downarrow}^n)^* + u_{j\downarrow}^n (v_{i\uparrow}^n)^*- u_{i\uparrow}^n (v_{j\downarrow}^n)^* - \nonumber
\\ u_{i\downarrow}^n (v_{j\uparrow}^n)^*\bigg] \tanh\bigg(\frac{E_n}{2k_{B}T}\bigg), \hspace{0.2cm} E_n > 0 \label{selfconsistent}
\end{align}
and is used to compute the energy eigenvalues and the  total energy.

In Fig.\ref{figfour},  we show the self-consistent energy eigenvalues in the presence of a 
vortex for a finite pairing interaction. We can compare these energy eigenvalues with those in Fig.\ref{figthree} without the pairing term
and note that there are four zero energy modes well separated from the bulk states, clearly showing that 
the system is a second order topological superconductor.  Moreover, the spectrum surprisingly shows a magnetic flux periodicity of 
$\frac{hc}{e}$ as opposed to a magnetic flux periodicity  of $\frac{hc}{2e}$ in a typical s-wave superconductor.

 The total self-consistent energy has been shown in Fig.\ref{figfive} for four different values of $b_x/\lambda$ as a  function of the flux $\phi/\phi_0$ through the centre of the lattice.
The total energy initially has a maximum at $\phi = \phi_0/2$ at $b_x=0$ which is in the topological superconductor regime as shown  in Fig.\ref{figfive}(a).  But as $b_x$ increases,
we note that the width of the maximum reduces (as in Fig.\ref{figfive}(b)), and then as shown in Fig.\ref{figfive}(c), the total energy develops
a minimum at $\phi=\phi_0/2$. Finally,  in Fig.\ref{figfive}(d), we note that at even higher values of $b_x$ where the model is no longer in 
the topological superconductor regime, the minimum at $\phi=\phi_0/2$ is again
replaced by a maximum. 
So as one increase $b_x$ from $b_x/\lambda=0$ , the magnetic flux periodicity of total energy which was $\phi_0$ to start with goes to $\phi_0/2$ at intermediate $b_x$ and then again goes back to $\phi_0$ as one further increase $b_x$.

Note that we have fixed the chemical potential at $\mu=0$ and have occupied two of the zero energy states - i.e., we are in the even parity state. As we change the flux, we keep the total number of particles fixed (the band has been precisely half-filled). 
So we are in the Coulomb blockaded regime where the parity of the state cannot change and continues to remain even.

The lattice current density $\vec{J}_{ij}$  
from lattice site $i$ to $j$
 is then obtained  as 
\begin{align}
\vec{J}_{ij}=-\frac{2e}{\hbar} \mathrm{I}\mathrm{m}\bigg[ {\vec t}_h \langle c_{j\uparrow}^{\dagger}c_{i\downarrow}\rangle \exp\big(i\phi_{ji}\big) - \nonumber 
\\ {\vec t}_h \langle c_{i\uparrow}^{\dagger}c_{j\downarrow}\rangle \exp\big(i\phi_{ij}\big) \bigg].
\label{current} 
\end{align}
using the continuity equation which is then used to compute the total circulating current
in the system, which is shown in Fig.\ref{figsix}. Here again, we note that as $b_x$ increases, there is a tendency towards period doubling $i.e$,   the flux periodicity changes from $\phi_0$ to $\phi_0/2$. Further increase in $b_x$ brings the periodicity again back to $\phi_0$, as shown in Fig.\ref{figsix}(c). Note however, that both the energy and the current density only show the tendency towards $\phi_0/2$ periodicity. It is not exactly $\phi_0/2$ periodic. This is because we have a finite system, with a small gap in the normal superconducting region,
and it is known that in the small gap regime, one does not get perfect $\phi_0/2$ periodicity\cite{Loder2012}.

\begin{figure}[h]
\centerline{
\begin{subfigure}{4.1cm}
\caption{}
\includegraphics[width=4.1cm]{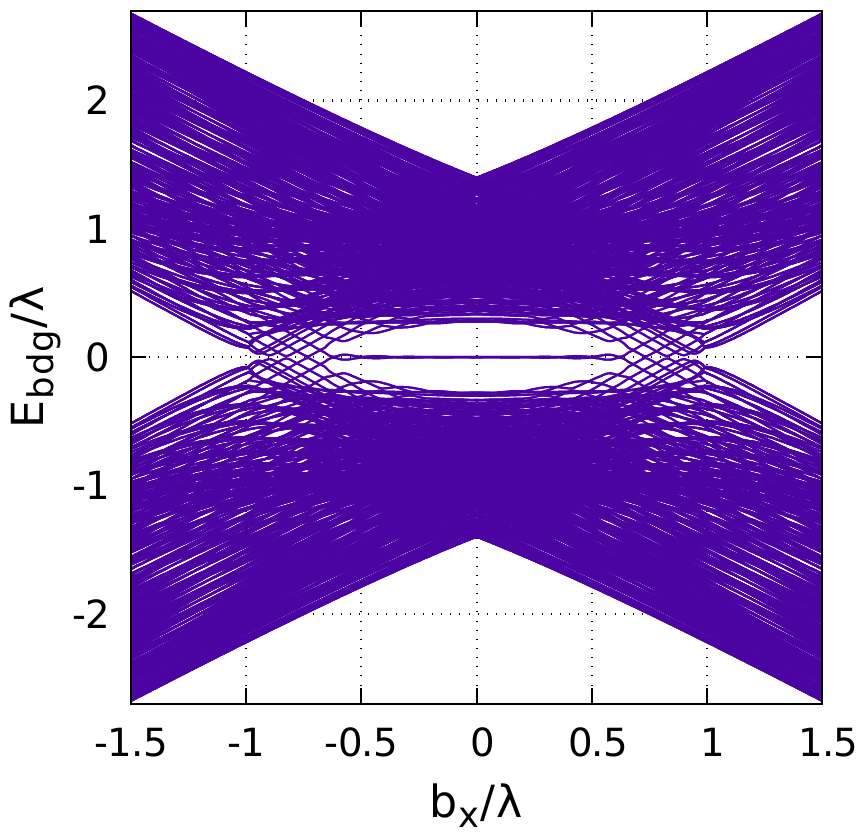}
\end{subfigure}
\begin{subfigure}{4.1cm}
\caption{}
\vspace{-0.1cm}
\includegraphics[width=4.0cm]{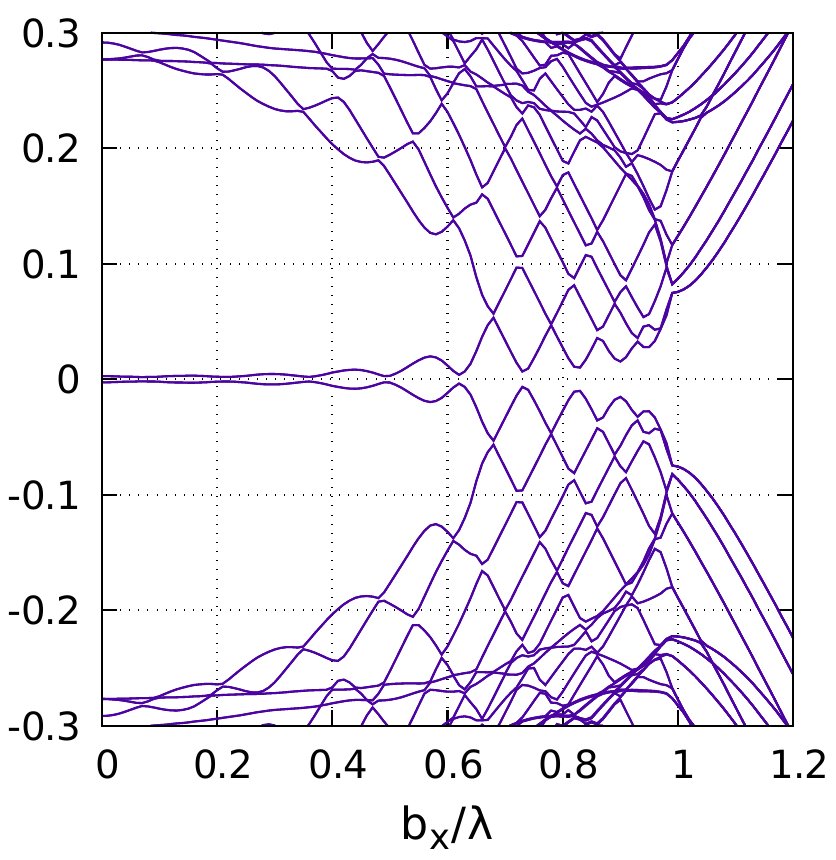}
\end{subfigure}
}
\caption{Self-consistent spectrum for the Hamiltonian in Eq.\ref{eqfive} on $20\times20$ square lattice having lattice spacing $a=1$, without any flux as function of field $b_x/\lambda$ for $\lambda=1,V=3\lambda$. (a)The full spectrum, (b) spectrum close to zero energy. These show that at close to $b_x/\lambda \simeq 0.6$ zero energy is gapped out by mixing with the bulk energy, giving an indication that there is a phase transition at that point.}
\label{figseven}
\end{figure}

Now to see why the magnetic flux periodicity of the system changes from $\phi_0$ to $\phi_0/2$ and goes back again to $\phi_0$,  we have calculated the self-consistent spectrum in the absence of the vortex but as a function of $b_x/\lambda$ as illustrated in Fig.\ref{figseven}.
The spectrum clearly shows  that  for small enough $b_x/\lambda$,  there are zero energy states well separated from the bulk states.  Here,  the system is in a second order topological superconductor phase and the magnetic flux periodicity of this topological phase is $\frac{hc}{e}$ as seen in the total energy in Fig.\ref{figfive}(a) and in  the total circulating current in  Fig.\ref{figsix}(a). Now close to $b_x/\lambda \simeq \pm 0.6$, the bulk energy gap closes and the zero energy states mix with bulk states giving rise to a continuum of energy states and signifying a change in  the  topology of the system.

\begin{figure}[h]
\centerline{
\includegraphics[width=7cm,height=4cm]{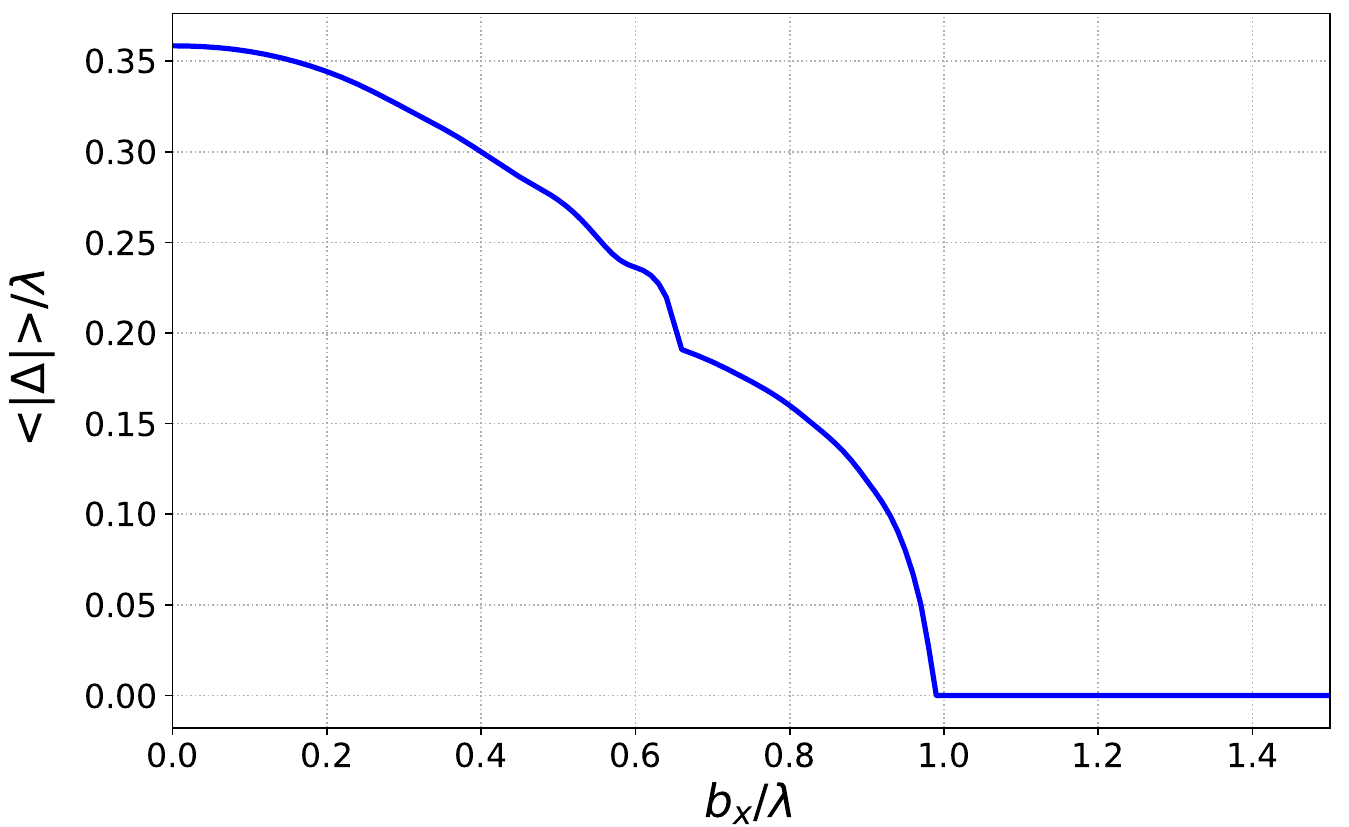}
}
\caption{Site average of the pairing term $\langle |\Delta|\rangle $ as a function of $b_x/\lambda$, calculated self-consistently without flux for the Hamiltonian in Eq.\ref{eqfive} on a $20\times20$ square lattice for $\lambda=1,V=3\lambda$ and lattice spacing $a=1$.  This clearly shows that the system is in a superconducting phase until $b_x \sim1.0$. Further increase of $b_x$ brings the system to a non-superconducting phase having zero pairing.}
\label{figeight}
\end{figure}

We have also calculated the site average of the  pairing term as a function of $b_x/\lambda$ as illustrated in Fig.\ref{figeight} to show that the system has finite pairing term upto $b_x/\lambda  \simeq 1.0$.
So the system remains  in a superconducting phase upto $b_x/\lambda \simeq 1.0$; however, close to $b_x/\lambda \simeq 0.6$ there  is a phase transition from a  topological to a non-topological superconductor, which can be seen from  the vanishing of the zero energy states. This causes the magnetic flux periodicity  to change from $\frac{hc}{e}$ to $\frac{hc}{2e}$ which is is seen both in  the  total energy and  in the total circulating current in  the system. So the magnetic flux periodicity change from $\frac{hc}{e}$ to $\frac{hc}{2e}$ is associated with the change of topology of the system, $i.e$, 2nd-order topological superconductor has flux-periodicity of $\frac{hc}{e}$ in contrast  to a non-topological superconductor which has  flux-periodicity of $\frac{hc}{2e}$.
When  $b_x$ is increased beyond $b_x/\lambda \simeq 1.0$,  the pairing term goes to zero and the system is in the metallic phase.
 This again explains the switch back  to the magnetic flux periodicity of $\frac{hc}{e}$ as seen both  in the total energy by Fig.\ref{figfive}(d) and in the total persistent current in  Fig.\ref{figsix}(c). This is just the  expected periodicity due to the  Aharanov-Bohm effect when a flux is inserted through  a metallic ring.

\begin{figure}[h]
\centerline{
\includegraphics[width=8.4cm,height=3.6cm]{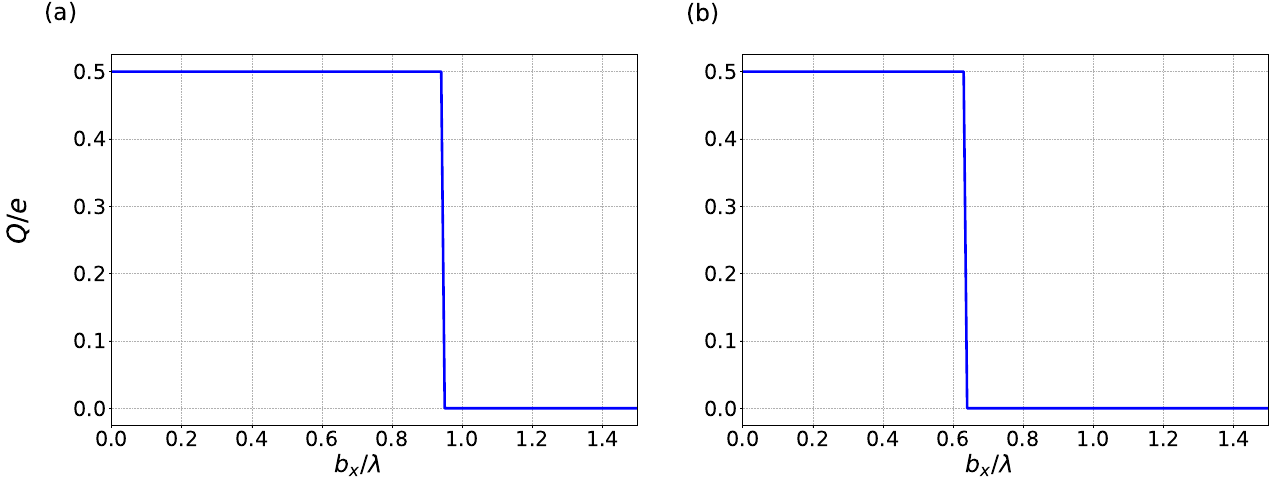}
}
\caption{The quadrupole moment($Q/e$) as a  function of the field $b_x/\lambda$ for (a) $\Delta/\lambda = 0.4$ (without self-consistency) for the Hamiltonian in Eq.\ref{eqone} and for (b) $ V = 3 \lambda $ (using self-consistency) for the Hamiltonian in Eq.\ref{eqfive} for $\lambda=1$, flux $\phi=0$ and for $20 \times 20$ square lattice having lattice spacing $a=1$ with open boundary condition in both $x$ and $y$ direction. In (b), the transition to a non-topological
phase occurs at  $b_x/\lambda \simeq 0.6$,  contrast  to (a) where the transition  happens at $b_x/\lambda \simeq 1.0$.}
\label{fignine}
\end{figure}

 We also confirm the phase transition from a topological superconductor to a normal superconductor at $b_x/\lambda \sim 0.6$ by calculating the
 the quadrupole moment($Q/e$) using the definition given in Eq.\ref{eqtwo} for the Hamiltonian in Eq.\ref{eqfive} without any flux. 
 As outlined in Ref.\cite{Agarwala2019}, we construct the  state $| \Phi_0 \rangle $ for the half filled system and then take the average of the operator $\hat{O}$ defined in Eq.\ref{eqtwo}.  However, here the filling refers to the filling of Boguliobons as defined in Eqs. \ref{eqseven} and \ref{eqeight}.
 As shown in Fig.\ref{fignine}(b) for low values of $b_x$, the system  is in the second order topological superconductor phase and the quadrupole moment
 is non-zero and given by $Q/e = 0.5$ (modulo 1). But at
   $b_x/\lambda \simeq 0.6$,  $Q/e$ become $0$ (modulo 1) indicating that the system transitions to the non-topological superconductor phase. 
   This is consistent with the spectrum which also shows a gap closing at the same point. 
This confirms that  the change in the magnetic flux periodicity from $\frac{hc}{e}$ to $\frac{hc}{2e}$ close to $b_x/\lambda \simeq 0.7$ for the system shown in Fig.\ref{figfive}(c) and Fig.\ref{figsix}(b) is associated with the change in the topology of the system.

\begin{figure}[h]
\centerline{
\includegraphics[width=8.4cm,height=7cm]{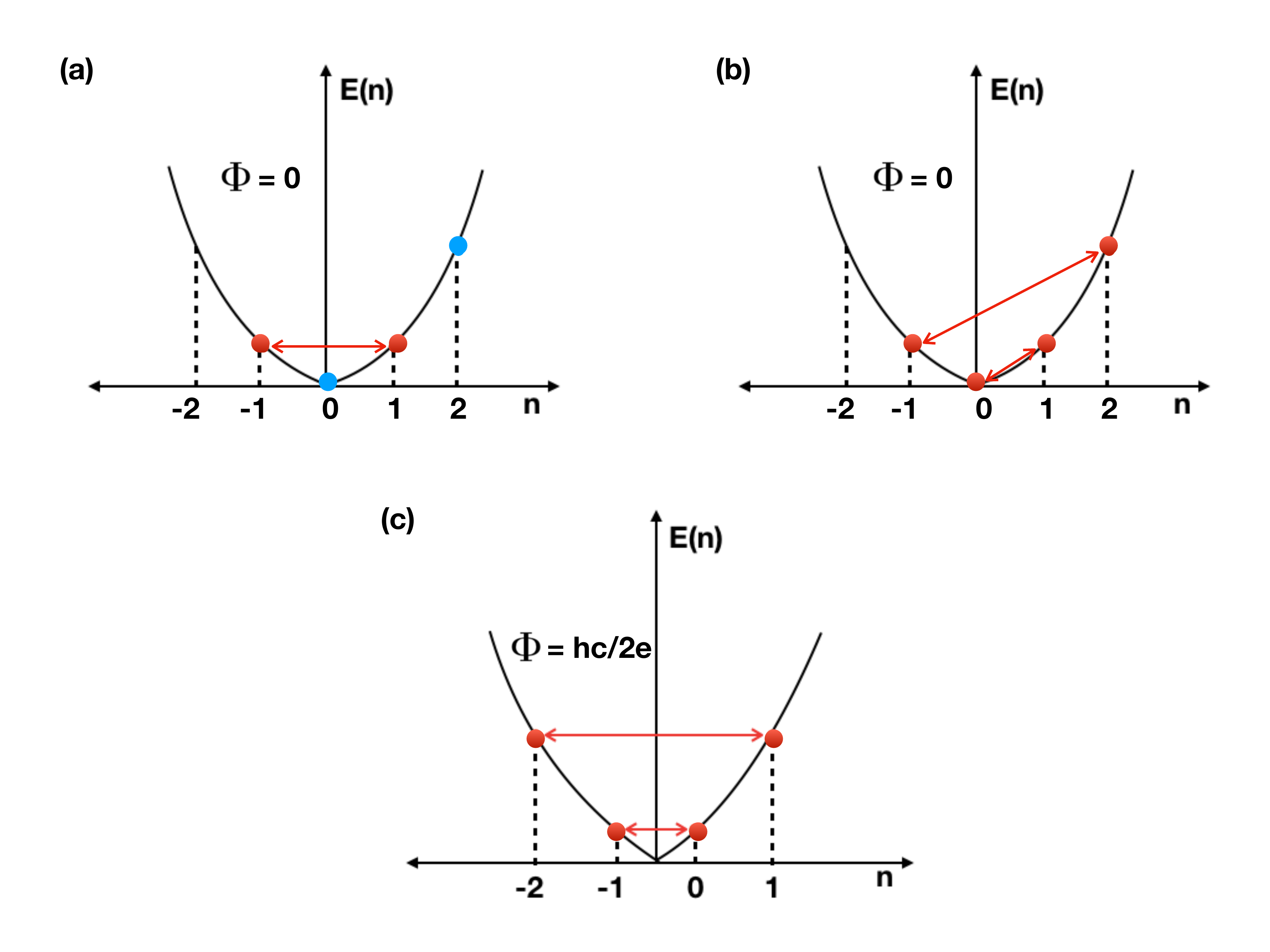}
}
\caption{One-particle energy levels of electrons on a ring. (a) Illustrates equal energy class I pairing at $\phi=0$, relevant
for weak pairing $p$-wave  topological superconductors. Note that the state at $\hbar k=0$ and $\hbar k=2$ are unpaired. (b) Illustrates
unequal energy class II  pairings relevant to strong coupling superconductors (for $b_x>0.6$). (c) Illustrates equal energy  class II pairings
at $\phi=\phi_0/2$, relevant for both weak and strong pairing $p$-wave cases. (The red dot with two-head arrow denotes paired electrons and the blue dot is for an unpaired electron) }
\label{fig:flux}
\end{figure}

Note that the change in periodicity can also be understood in a simple one-particle picture in the following way. As mentioned earlier in the section on vortex insertion in an annulus geometry, for normal $s$-wave superconductors, there are two classes of pairing wave-functions - the class I wave-functions where the pairing is between $\hbar (+k)$ and $\hbar(-k)$ at $\phi=0$ and the class II pairing wave-functions where the pairing
is between $\hbar(+k)$ and $\hbar(-k+1)$ at $\phi=\phi_0/2$.  The pairing energy is exactly the same for both these classes of wave-functions
and so there is perfect degeneracy between the ground state energy at $\phi=0$ and at $\phi=\phi_0/2$.  This is what leads
to the $\phi_0/2$ periodicity of a normal $s$-wave superconducting ring. However, for a weak pairing $p$-wave superconductor, the orbital antisymmetry does not allow  two electrons in the $k=0$ state at $\phi=0$.  Thus, there are unpaired electrons at $k=0$ and at the Fermi energy, so one pair of electrons remains
unpaired, as shown in Fig.\ref{fig:flux}(a).  However, at $\phi=\phi_0/2$, the energy minimum shifts and the  pairings are now the class II pairings, which  allows for
equal energy pairings  for
all values of momenta.  This is illustrated in Fig.\ref{fig:flux}(c). This breaks the exact degeneracy between the ground states at $\phi=0$ and $\phi=\phi_0/2$.
Thus for $p$-wave topological superconductors,which is weak pairing $p$-wave superconductor the periodicity is $\phi_0$ and not $\phi_0/2$.

However, for $b_x>0.6$, where our results show that there is a phase transition from topological to non-topological superconductor,
it is likely that the transition is to the strong pairing $p$-wave phase where the pairing can occur  even between states that are slightly different in energy,
as shown in Fig.\ref{fig:flux}(b).
In that case, even at $\phi=0$, all states are paired and the degeneracy between the ground states at $\phi=0$ and $\phi=\phi_0/2$ is restored.

We also note  that a similar result of period doubling in the semiconductor-superconductor nanowires with $p+ip$ superconductivity was earlier studied by Zocher {\it et al}\cite{Zocher2012,Zocher2013}.  For even parity, they also showed that
for fixed mean particle number, the excitation spectrum showed clear signatures of the $\frac{hc}{e}$ periodicity in the topologically non-trivial phase.

\section{Discussions and conclusions}

In this work, we have focused on the magnetic flux periodicity of a second order topological superconductor in two 
dimensions with four Majorana modes at the corners. By implementing a ring geometry via a flux at the origin, we show 
that the flux periodicity changes from $\frac{hc}{e}$ to $\frac{hc}{2e}$ when the topological superconductor transitions 
to an ordinary superconductor. 

This model hosts four Majorana modes at its corners, which is sufficient to exploit their non-abelian nature in braiding 
since they can be paired in different ways~\cite{Beenakker2019}. By insertion of multiple vortices and generalizing our 
model, one can expect to braid the Majorana modes through adiabatic manipulations of fluxes. This could be useful in 
designing non-abelian quantum computational protocols. Therefore higher order topological insulators could open a new pathway 
towards topological quantum computation.

\begin{acknowledgments}

U.K. was supported by the Raymond and Beverly Sackler Faculty of Exact Sciences at Tel Aviv University and the Raymond 
and Beverly Sackler Center for Computational Molecular and Material Science. We thank A. Agarwala, S. Kadge, G. Murthy, 
B. Rosenow and Y. Gefen for useful discussions.

\end{acknowledgments}

\end{document}